\documentclass[a4paper,11pt]{article}
    \usepackage{aaskaiid}
    \usepackage{hyperref}
    \usepackage{enumitem}
    \usepackage{wrapfig}
    \usepackage{orcidlink}
    \def\hi{H\,{\sc i}}
    \def\hii{H\,{\sc ii}}
    \title{SKAO and Gamma-Ray Synergies}
    \ShortTitle{SKAO and Gamma-Ray Synergies}

    \author[1]{Gianluca Castignani$^\dagger$\orcidlink{0000-0001-6831-0687}}
    \author[2]{Gavin Rowell$^\dagger$\orcidlink{0000-0002-9516-1581}}
    \ShortName{Castignani, Rowell et al.} 
    \author[3]{Arnau Aguasca-Cabot\orcidlink{0000-0001-8816-4920}}
    \author[4,5]{Gemma E. Anderson\orcidlink{0000-0001-6544-8007}}
    \author[6]{Csaba Balazs\orcidlink{0000-0001-7154-1726}}
    \author[3]{Pol Bordas\orcidlink{0000-0002-0266-8536}}
    \author[7]{Andrea Botteon\orcidlink{0000-0002-9325-1567}}
    \author[8]{Jess W. Broderick\orcidlink{0000-0002-2239-6099}}
    \author[7]{Gianfranco Brunetti\orcidlink{0000-0003-4195-8613}}
    \author[7]{Ettore Carretti\orcidlink{0000-0002-3973-8403}}
    \author[9]{Roland M.~Crocker\orcidlink{0000-0002-2036-2426}}
    \author[4]{Shi Dai\orcidlink{0000-0002-9618-2499}}
    \author[7]{Filippo D'Ammando\orcidlink{0000-0001-7618-7527}}
    \author[4]{Philip G. Edwards\orcidlink{0000-0002-8186-4753}}
    \author[2]{Sabrina Einecke\orcidlink{0000-0001-9687-8237}}
    \author[10]{Miroslav D. Filipovi\'c\orcidlink{0000-0002-4990-9288}}
    \author[7]{Marcello Giroletti\orcidlink{0000-0002-8657-8852}}
    \author[11]{Adelle J. Goodwin\orcidlink{0000-0003-3441-8299}}
    \author[8]{James A. Green\orcidlink{0000-0002-2670-188X}}
    \author[10,4]{Sanja Lazarevi\'c\orcidlink{0000-0001-6109-8548}}
    \author[7]{Giulia Migliori\orcidlink{0000-0003-0216-8053}}
    \author[7]{Monica Orienti\orcidlink{0000-0003-4470-7094}}
    \author[3]{Josep M. Paredes\orcidlink{0000-0002-1566-9}}
    \author[1]{Elena Pian\orcidlink{0000-0001-8646-4858}}
    \author[12]{Alak Ray\orcidlink{0000-0003-2404-0018}}
    \author[13,14]{Lauren Rhodes\orcidlink{0000-0003-2705-4941}}
    \author[10]{Zachary J. Smeaton\orcidlink{0009-0009-7061-0553}}
    \author[10]{Nick Tothill\orcidlink{0000-0002-9931-5162}}
    \author[15]{Martin White\orcidlink{0000-0001-5474-4580}}
    \emailAdd{gianluca.castignani@inaf.it}
    \emailAdd{gavin.rowell@adelaide.edu.au}

    \affiliation[1]{INAF - Osservatorio di Astrofisica e Scienza dello Spazio di Bologna, via Gobetti 93/3, I-40129, Bologna, Italy}

    \affiliation[2]{School of Chemistry, Physics \& Earth Sciences, The University of Adelaide, Adelaide 5005, Australia}

    \affiliation[3]{Departament de Física Quàntica i Astrofísica, Institut de Ciències del Cosmos, Universitat de Barcelona, IEEC-UB, Martí i Franquès 1, 08028 Barcelona, Spain}

    \affiliation[4]{Australia Telescope National Facility, CSIRO, Space and Astronomy, P.O. Box 76, Epping, NSW 1710, Australia}
    
    \affiliation[5]{Sydney Institute for Astronomy, School of Physics, The University of Sydney, NSW 2006, Australia}
   
    \affiliation[6]{School of Physics and Astronomy, Monash University, Melbourne 3800 Victoria, Australia}

    \affiliation[7]{INAF - Istituto di Radioastronomia, Via Gobetti 101, I-40129 Bologna, Italy}

     \affiliation[8]{SKA Observatory, Science Operations Centre, CSIRO ARRC, 26 Dick Perry Avenue, Kensington, WA 6151, Australia}

    \affiliation[9]{Research School of Astronomy and Astrophysics, Australian National University, Weston 2611, Australia}

    \affiliation[10]{Western Sydney University, Locked Bag 1797, Penrith South DC, NSW 2751, Australia}

    \affiliation[11]{International Centre for Radio Astronomy Research, {Curtin University}, {GPO Box U1987}, {Perth}, {6845}, {WA}, {Australia}}

    \affiliation[12]{Tata Institute of Fundamental Research, Mumbai 400005, India}

    \affiliation[13]{Trottier Space Institute, McGill University, 3550 rue University, Montr\'eal, QC H3A 2A7, Canada}
    
    \affiliation[14]{Department of Physics, McGill University, 3600 rue University, Montr\'eal, QC H3A 2T8, Canada}

    \affiliation[15]{University of Adelaide, ARC Centre of Excellence for Dark Matter Particle Physics \& CSSM, Department of Physics, Adelaide SA 500 Australia \\\vspace{0.5cm}  $^\dagger$ These authors contributed equally to this work and are corresponding authors.}

\abstract{

{\bf Abstract.} A wide variety of Galactic and extragalactic sources are known to emit radiation across the entire electromagnetic spectrum, including both transient and steady-state phenomena. A few hundred of these sources ($\sim300$) have been detected even at the highest energies, in the TeV range. The number of known TeV emitters is expected to increase substantially in the coming years with the operation of current and next-generation Cherenkov detectors, such as the Large High Altitude Air Shower Observatory (LHAASO) and the Cherenkov Telescope Array Observatory (CTAO).  These sources typically exhibit broad, non-thermal, spectral energy distributions. Explaining such emission requires efficient particle acceleration mechanisms (e.g. Fermi processes, shock acceleration) and radiative processes involving magnetic fields (e.g. synchrotron and inverse Compton radiation), often accompanied by polarization signatures. However, the relative contribution of these emission mechanisms and the underlying physical processes are still debated. In this work, we present an overview of the scientific potential arising from the synergy between the Square Kilometre Array (SKA) and current and upcoming gamma-ray facilities. Combined observations across these energy bands will provide crucial insights into the physical mechanisms driving emission from GeV–TeV sources of both Galactic and extragalactic origin. These include transient events (e.g. gamma-ray bursts, supernovae, fast radio bursts, tidal disruption events, neutrino and gravitational-wave counterparts), variable sources (e.g. blazars, active galactic nuclei), and steady emitters (e.g. the Galactic centre, supernova remnants, radio galaxies, and galaxy clusters). We discuss the prospects for coordinated SKA–gamma-ray observations, including wide-field surveys, monitoring of variable sources, and target-of-opportunity follow-ups, emphasizing their role in advancing our understanding of high-energy astrophysical processes.}

\begin{document}
    \newcommand{\actaa}{Acta Astron.} 
\newcommand{\araa}{Annu. Rev. Astron. Astrophys.} 
\newcommand{\aar}{Astron. Astrophys. Rev.} 
\newcommand{\ab}{Astrobiol.} 
\newcommand{\aj}{Astron. J.} 
\newcommand{\apj}{Astrophys. J.} 
\newcommand{\apjl}{Astrophys. J. Lett.} 
\newcommand{\apjs}{Astrophys. J. Suppl. Ser.} 
\newcommand{\ao}{Appl. Opt.} 
\newcommand{\apss}{Astrophys. Space Sci.} 
\newcommand{\aap}{Astron. Astrophys.} 
\newcommand{\aapr}{Astron. Astrophys. Rev.} 
\newcommand{\aaps}{Astron. Astrophys. Suppl.} 
\newcommand{\baas}{Bull. Am. Astron. Soc.} 
\newcommand{\caa}{Chinese Astron. Astrophys.} 
\newcommand{\cjaa}{Chinese J. Astron. Astrophys.} 
\newcommand{\cqg}{Class. Quantum Gravity} 
\newcommand{\gal}{Galaxies} 
\newcommand{\gca}{Geochim. Cosmochim. Acta} 
\newcommand{\icarus}{Icarus} 
\newcommand{\jcap}{J. Cosmol. Astropart. Phys.} 
\newcommand{\jgr}{J. Geophys. Res.} 
\newcommand{\jgrp}{J. Geophys. Res.: Planets} 
\newcommand{\jqsrt}{J. Quant. Spectrosc. Radiat. Transf.} 
\newcommand{\memsai}{Mem. Soc. Astron. Italiana} 
\newcommand{\mnras}{Mon. Not. R. Astron. Soc.} 
\newcommand{\nat}{Nature} 
\newcommand{\nastro}{Nat. Astron.} 
\newcommand{\ncomms}{Nat. Commun.} 
\newcommand{\nphys}{Nat. Phys.} 
\newcommand{\na}{New Astron.} 
\newcommand{\nar}{New Astron. Rev.} 
\newcommand{\physrep}{Phys. Rep.} 
\newcommand{\pra}{Phys. Rev. A} 
\newcommand{\prb}{Phys. Rev. B} 
\newcommand{\prc}{Phys. Rev. C} 
\newcommand{\prd}{Phys. Rev. D} 
\newcommand{\pre}{Phys. Rev. E} 
\newcommand{\prl}{Phys. Rev. Lett.} 
\newcommand{\psj}{Planet. Sci. J.} 
\newcommand{\planss}{Planet. Space Sci.} 
\newcommand{\pnas}{Proc. Natl Acad. Sci. USA} 
\newcommand{\procspie}{Proc. SPIE} 
\newcommand{\pasa}{Publ. Astron. Soc. Aust.} 
\newcommand{\pasj}{Publ. Astron. Soc. Jpn} 
\newcommand{\pasp}{Publ. Astron. Soc. Pac.} 
\newcommand{\rmxaa}{Rev. Mexicana Astron. Astrofis.} 
\newcommand{\sci}{Science} 
\newcommand{\sciadv}{Sci. Adv.} 
\newcommand{\solphys}{Sol. Phys.} 
\newcommand{\sovast}{Soviet Ast.} 
\newcommand{\ssr}{Space Sci. Rev.} 
\newcommand{\uni}{Universe} 

    \maketitle
    
    
    \section{Introduction}
    \label{Intro}

    This chapter looks at some of the linkages across both radio and gamma-ray astronomy, which represents the extreme ends of the electromagnetic spectrum. With a number of next-generation gamma-ray facilities commencing within the SKAO era, it is important to review a number of key scientific challenges that the SKAO and these new facilities are expected to tackle.

    The radio and gamma-ray connection is underpinned by production mechanisms powered by high-energy (relativistic) charged particles. Such particles (electrons, protons, nuclei) can be accelerated in extreme environments typically associated with a variety of astrophysical scenarios, leading to a non-thermal distribution of particle energies.
    
    The scenarios include supernovae and their remnants, compact objects (white dwarfs, neutron stars and black holes) and their surroundings, collimated jets and outflows associated with accretion onto compact objects, compact object mergers, and stellar wind interactions in stars and stellar clusters. Radio and gamma-ray emission can also play critical roles as potentially detectable signatures of beyond-standard-model physics, such as in models of dark matter and in models of other 'relics' from the early Universe. 

    The key production processes are  \citep[see][for more detailed reviews]{Ahar:2004,Hinton:2009}:
    \begin{itemize}[leftmargin=1em]
        \item {\em Synchrotron emission}: This is produced by particles propagating in magnetic fields. For typical magnetic fields, in the nG (intergalactic space) to G range (inside jets), radio synchrotron emission is dominantly produced by electrons with GeV energies. Related X-ray synchrotron emission also arises when electrons reach TeV energies. For extreme magnetic fields $>10^{11}$G around neutron stars (pulsars, magnetars), particles are tightly constrained to magnetic field lines, leading to curvature radiation as the source of radio emission.
       \item {\em Inverse-Compton (IC) scattering}: These same electrons can also produce IC gamma-ray emission from GeV to potentially PeV energies. Here, electrons can up-scatter low-energy photon fields such as the cosmic microwave background and local infrared/optical fields. The IC and synchrotron emissions are thus tightly coupled and their relative strengths are governed by the energy densities of the magnetic field and that of the low-energy photon fields \citep{Aharonian:1997}. To first order, the spatial distribution of the synchrotron and IC emission can be somewhat anti-correlated, as synchrotron emission dominates in high-magnetic-field regions, in contrast to that of IC emission (since the two processes compete for the energy of the same electrons).
       \item{\em Proton-proton collisions}: Cosmic-ray (CR) protons of at least GeV energies can interact with ambient medium (molecular, atomic or ionised gas) to produce gamma-rays in the GeV to PeV energy range (via the decay of $\pi^\circ$ particles). Spatial correlation between the ambient gas and the resulting gamma-ray emission is expected in this case, although complications occur if low-energy CRs ($<1$\,TeV) cannot penetrate dense gas  \citep[e.g.][]{Gabici:2007}. Electrons can also be produced as a byproduct of the CR interaction (resulting from the decay chain $\pi^\pm \rightarrow \mu^\pm \rightarrow e^\pm$), leading to a secondary synchrotron emission component often in the radio band.
    \end{itemize}
    
In terms of gamma-ray astronomy, our focus is on the energy range of about 0.1\,GeV and above. For these energies, the above-listed processes are usually the dominant energy loss mechanisms for energetic particles. The key gamma-ray facilities used to detect such gamma-rays today include the space-based {\em Fermi}-LAT, and the ground-based H.E.S.S.\ \citep{HESS}, MAGIC \citep{MAGIC}, VERITAS \citep{VERITAS}, Tibet AS-$\gamma$ \citep{Tibet}, ARGO-YBJ \citep{ARGO}, HAWC \citep{HAWC} and LHAASO \citep{LHAASO}. All of these have built on the success of earlier-generation facilities using either direct detection (space-based) or the detection of secondary particles and Cherenkov photons generated when gamma rays interact with the Earth's atmosphere (ground-based). The reader is referred to \cite{Fegan:2019,Chadwick:2021,Sitarek:2022} for historical reviews of ground-based gamma-ray detection methods and facilities.

\begin{wrapfigure}{r}{0.7\textwidth}
    \centering
    \includegraphics[width=0.7\textwidth]{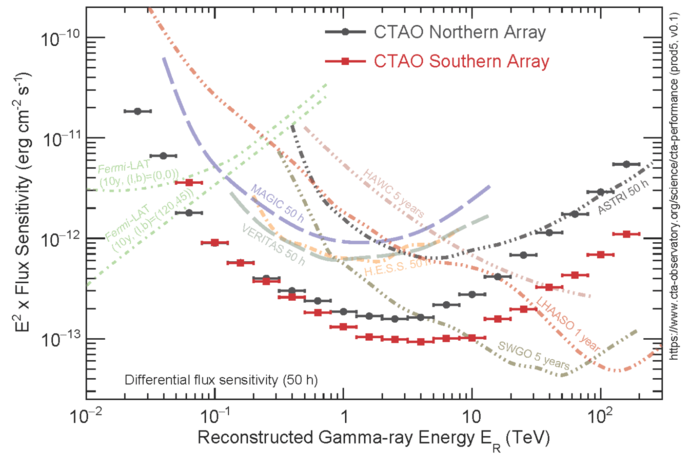}
    \caption{Energy-flux sensitivity of CTAO's North and South arrays along with comparisons to various other facilities. Image taken from https://www.ctao.org/for-scientists/performance/}
    \label{fig:CTAOsens}
\end{wrapfigure}

The next-generation in ground-based gamma-ray astronomy, aiming for $>$10x improvement in sensitivity (see Fig.\,\ref{fig:CTAOsens}) is focusing on expanded arrays of Cherenkov telescopes and/or particle detectors. The Cherenkov Telescope Array Observatory (CTAO) will operate both Northern (La Palma) and Southern (Paranal, Chile) sites hosting 13 and 58 Cherenkov telescopes respectively \citep{CTA}. The Southern Wide Field Gamma-ray Observatory (SWGO), at a high altitude site in Chile (4770\,m), aims for a dense array of particle detectors spread over a $\sim$1\,km$^2$ area \citep{SWGO}. Additional arrays of Cherenkov telescopes are also planned for the Northern hemisphere -- ASTRI \citep{ASTRI} at Tenerife, and LACT \citep{LACT} in China.
CTAO and SWGO are complementary with CTAO's high sensitivity and arc-min angular resolution balanced by SWGO's 24\,hr operation and wide $\sim \pi$\,sr field of view. Importantly for the SKAO, both of these new facilities will be situated in the Southern Hemisphere, thus providing a unique opportunity to link in with SKAO's science. {We also note the potential of the SKAO itself in detecting PeV gamma-rays \citep{Nelles01.2026.SKA}}.
We describe below a number of possible scientific cases for the SKAO. These include a broad variety of radio and gamma-ray emitting source types, both Galactic and extra-galactic, to be observed by the SKAO as targets of opportunity, targeted observations, and/or observation via survey mode. The cases are ordered from small to large-scale based on the general sizes of the regions responsible for radio and gamma-ray emission.





\section{Variable and Transient Sources}\label{sec:variable_and_transients}

   \subsection{Pulsars, Magnetars, and Fast Radio Bursts}

Identifying a large and diverse sample of gamma-ray pulsars enables robust tests of pulsar emission models by comparing predicted and observed properties with minimal selection bias~\citep[e.g.][]{2023ApJ...958..191S,2025OJAp....854244O}. It also improves population extrapolations, helping to estimate the contribution of unresolved pulsars to the diffuse gamma-ray background~\citep[e.g.][]{2021PhRvD.103f3029D}.
Pulsed very-high-energy gamma rays (above 100\,GeV) have been observed from the most energetic pulsars in the Milky Way, providing crucial insights into acceleration and emission processes at their extreme limits~\citep{2016A&A...585A.133A,2023NatAs...7.1341H}.
For known radio pulsars, gamma-ray pulsation can be effectively searched by folding gamma-ray data with phase-connected rotation ephemerides that remain accurate over the full span of the observations~\citep[e.g.][]{2019ApJ...871...78S}. Long-term radio monitoring of a large pulsar sample provides the required timing measurements for such precise rotation ephemerides~\citep[e.g.][]{2010PASA...27...64W,2021MNRAS.502.1253J}.

New pulsars can be discovered through blind periodicity searches in gamma-ray data sets, which do not require prior knowledge of their spin periods or spin-down rates~\citep[e.g.][]{2009Sci...325..840A,2017ApJ...834..106C}. This has been made possible largely thanks to powerful gamma-ray instruments such as the Large Area Telescope (LAT) aboard the Fermi Gamma-ray Space Telescope~\citep{2009ApJ...697.1071A}. An even more effective approach is conducting targeted radio searches of Fermi-LAT unassociated sources~\citep[e.g.][]{2012arXiv1205.3089R}. Such targeted searches enable deep observations with large radio telescopes~\citep[e.g.][]{2011ApJ...727L..16R,2011MNRAS.414.1292K,2012ApJ...748L...2K,2013MNRAS.429.1633B,2016ApJ...819...34C,2017ApJ...846L..19P,2021ApJ...910..160B,2021SCPMA..6429562W,2023MNRAS.519.5590C,2024ApJ...966..161B} or wide-band observing systems~\citep{2020ApJ...888L..18D}, and the use of advanced algorithms to identify pulsars in compact binary systems~\citep{2021cosp...43E1208T}. 
{These efforts have led to the discovery of over one quarter of the approximately 400 millisecond pulsars  known in the Galactic field, and we anticipate that further discoveries of millisecond pulsars will be enabled by targeted searches with SKA-Mid and SKA-Low~\citep{2025OJAp....854256K,2025OJAp....854252A,2025OJAp....854251B,Bagchi01.2026.SKA,Abbate01.2026.SKA,Keane01.2026.SKA}.}

Timing pulsars using gamma-ray data complements radio observations. Unlike radio wavelengths, gamma-rays are unaffected by propagation effects caused by the ionised interstellar medium or the solar wind -- factors that limit the precision of radio timing. Using 12.5 years of Fermi-LAT data from the 35 brightest and most stable gamma-ray MSPs, \citet{2022Sci...376..521F} searched for a gravitational wave background (GWB) and placed upper limits on its amplitude. This direct measurement provides an independent probe of the GWB and serves as a valuable cross-check for noise models derived from radio observations.

So far, no gamma-ray emission ($\gtrsim1$\,GeV) has been detected directly from known magnetars~\citep{2017ApJ...835...30L,2025JCAP...07..050R}, a rare class of neutron stars with extremely strong magnetic fields ($\gtrsim10^{14}$\,G). 
{The non-detection of VHE emission may indicate that nonthermal processes (e.g. inverse Compton scattering) are suppressed in the magnetar environment. One possibility is that gamma-ray production occurs very close to the magnetar surface, where pair production and photon splitting lead to substantial energy losses, producing strong cutoffs in the MeV--GeV range~\citep[e.g.][]{Abdalla2021}. Deep observations with next generation Gamma-ray telescopes will allow us to test these scenarios. Conversely, any detection of VHE emission from magnetars would point to emission regions located well beyond the influence of the magnetar’s intense magnetic field~\citep[e.g.][]{2019MNRAS.486.3327H}.}
Radio emission has been observed from only 6 of the roughly 30 known magnetars, and these sources exhibit highly variable and diverse properties at radio wavelength~\citep[e.g.][]{2019ApJ...874L..14D,2019ApJ...882L...9M,2020MNRAS.498.6044C,2021MNRAS.502..127L,2023ApJ...945..153L}. In the case of the magnetar SGR 1935+2154, intense radio pulses were detected simultaneously to X-ray bursts in April 2020 \citep{Bochenek2020,CHIME/FRB2020,Mereghetti2020}. 
Detecting radio and gamma-ray emission from magnetars, or in association with their X-ray flares, is crucial for understanding the radiation mechanisms that power these objects~\citep[see][for a review]{2022ARA&A..60..495P}; however, the lack of any TeV detection by HESS in association with the bursts from SGR\,1935+2154 suggests that absorption of the TeV gammas may be at work \citep{Abdalla2021}. 

SGR\,1935+2154 is also considered the only galactic fast radio burst~\citep[FRB, see][for a review]{2023RvMP...95c5005Z}, with the rest of the population being found in other galaxies, out to cosmological distances. The SKAO will play a crucial role in detecting and precisely locating thousands of FRBs, contributing to mapping cosmic matter and better understanding the origins and physics behind these mysterious signals \citep{Caleb01.2026.SKA,Caleb02.2026.SKA,Curtin01.2026.SKA}. 
{To date, no high-energy bursts have been detected from known FRBs, despite this being a key objective for FRB research in the coming years.} Several models predict gamma-ray emission from FRBs \citep[e.g.][]{2019MNRAS.485.4091M}, although the sensitivity of the present instrumentation has only provided upper limits \citep{Principe2023}. {Deep gamma-ray follow-up observations of FRBs, particularly repeaters, with next-generation telescopes will be crucial for uncovering the origin of these enigmatic sources.}



    \subsection{Gamma-Ray Bursts}\label{sec:GRBs}

Gamma-ray bursts (GRBs) are brief, bright flashes of $\sim$keV to MeV gamma-rays that are produced when highly relativistic jets are launched in cataclysmic explosions. 
The distribution of GRB durations is bimodal, with a separation around 2 seconds between short-duration spectrally-hard bursts, and long-duration spectrally-soft bursts \citep{Kouveliotou1993}.  
Long GRBs, which make up $\sim$75\% of the population, are associated with massive stars undergoing core-collapse supernovae, referred to as collapsars \citep{Galama98,Woosley1999A,Hjorth03,stanek03}, whereas short GRBs arise from compact object binary mergers \citep[neutron star--neutron star or neutron star--black hole,][]{Ruffert99,Rosswog03,Hotokezaka11,abbott17b}.

Following the prompt flash of gamma-ray bursts is a longer lasting, broadband emission signature called an `\textit{afterglow}'. The afterglow spans from radio to TeV energies and originates from external shocks between the jet and its environment, which produce synchrotron emission \citep{piran99,kobayashi00sari} and, possibly, inverse-Compton radiation, either via synchrotron self-Compton (SSC) or external inverse-Compton from seed electrons \citep{Sari_2001}.
The external shocks are usually modelled under the forward-reverse shock framework, where the forward shock is generated by the jet propagating into the surround medium, which creates a short-lived reverse shock that moves back into the post-shocked medium \citep{meszaros97}. 
Given that synchrotron emission models predict a maximum possible photon energy of $\sim10$\,GeV \citep{piran10}, the likely source of higher energy (e.g. TeV) photons in the afterglow is from inverse Compton emission.

At radio frequencies, multi-frequency, multi-epoch observations are used to track the emission from different shock components that constitute the synchrotron afterglow \citep{granot02, granot14}. 
By understanding the locations of the different synchrotron break frequencies from the different shock components it is possible to use the radio emission (in combination with optical and X-ray data) to extract information on the 
energetics, circumburst environment and jet microphysics \citep{Meszaros_1998, Sari1998, Chevalier1999}. The models used to derive the aforementioned jet parameters, rely on knowing the hydrodynamics of the jet. Radio observations are vital in modelling and understanding GRB afterglows for, without radio observations, the physical parameters extracted from any modelling attempt will be highly degenerate.
Radio observations are also the main wavelength for detecting multiple synchrotron emission components. The reverse shock component evolves rapidly (seconds to minutes) at optical and X-ray wavelengths, but evolves on hour to $\sim1$\,day timescales in the radio band \citep[e.g.][]{anderson14,anderson24,vanderhorst14}.
Meanwhile, late time radio detections, including VLBI imaging for suitable GRBs, can reveal multiple forward shock components, indicating structured outflows \citep[e.g.][]{2018Natur.554..207M,Mooley2018,rhodes22,Rhodes2024}.



At the time of writing, there are only seven GRBs with detections at very high energies (VHE, $>$100\,GeV): six long GRBs and one short. By comparison, $\sim150$ bursts have been detected by Fermi-LAT at $>20$\,MeV \citep{Abdalla2019, magic_2019, HESS_2021, MAGIC_2021, Suda2022,Cao2023, abe_2024}. Each VHE event was also detected at radio frequencies \citep{lamb_2019, Rhodes2020, Misra_2021, Giarratana2022, Rhodes_2022, bright2023}. 
However, in some cases, the detection of TeV photons from GRBs has called into question the inverse Compton (SSC) interpretation of this high-energy component. \citet{HESS_2021} modelled H.E.S.S., Fermi-LAT and X-ray data from Swift \citep{swift,burrows05} from GRB~190829A to demonstrate that SSC could not describe the optical to TeV spectrum but was best fit with a single synchrotron origin, violating the maximum synchrotron photon energy assumptions.
The observation of very high-energy afterglow emission from GRB~221009A with LHAASO \citep{Cao2023} provides an exceptional probe of particle acceleration in relativistic outflows. The hard multi-TeV spectrum, however, challenges the synchrotron self-Compton interpretation of the conventional one-zone afterglow model \citep[e.g.][]{Ren_2024,Zhang2024,Zheng_2024}

One such possibility is the detected TeV emission from GRBs is a superposition of SSC from multiple synchrotron components, rather than just a single forward shock, which is usually assumed. 
It is therefore only by tracking the much slower evolving radio emission from $<1$\,day to late times (up to hundreds of days) can we decompose the multi-wavelength afterglow into its (potentially multiple) reverse and forward shock components, and resolve late-time outflow features (e.g. cocoons) to gain insight into the true TeV emission mechanism(s) \citep[e.g.][]{Rhodes2024}. 
This also requires rapid-response (automated) radio follow-up of GRBs to be on target within hours post-burst to track the reverse shock evolution within $\sim1$\,day post-burst \citep[e.g.][]{anderson14,anderson24} but which can also detect unexpected or additional early time emission from interstellar scintillation \citep{anderson23} or from collisions within the early outflow \citep{anderson25pp}.  
Indeed, detecting afterglow emission over such a broad frequency range has enabled the most complete modelling of GRBs to date and expanded our understanding of their total energy budget and the particle acceleration processes within the jet \citep[e.g. ][]{Klinger_2023, Barnard_2024}.

Despite the advancements in the area of VHE GRBs, as a community we are still limited by the low number of events. In order to continue to progress, we need more VHE GRBs with radio detections where the radio detections are earlier enough post burst that they are simultaneous with the VHE counterparts only possible with rapid response radio observations. The superior sensitivity of SKA-Low and SKA-Mid will enable them to detect fainter afterglows, monitor them over longer timescales, and anchor VLBI follow-up \citep{Giarratana01.2026.SKA}. A rapid response capability would be of great benefit in constraining the poorly understood, earlier time radio counterparts.

\subsection{High-Energy Core-Collapse Supernovae (SNe)}

{The interaction of a SN explosion blast with the circumstellar medium causes diffusive shock acceleration of particles that produce radiation, via bremsstrahlung or non-thermal processes,  over the whole electromagnetic spectrum from radio wavelengths to TeV energies, and may escape at relativistic energies as cosmic rays.  In particular, protons accelerated at the high-speed SN shock (up to $\sim$20000~km~s$^{-1}$) can interact with dense circumstellar material or photon fields to produce secondary pions, which decay into very high-energy photons that can be detected with CTAO.
Theoretical models link radio emission to shock dynamics, magnetic field amplification, and particle acceleration efficiency \citep{Chakraborti_Ray2011,Chakraborti2011NatCo,Chakraborti2015,BarniolDuran_Giannios2015}.} 
{While in young SNe the near-infrared to UV part of the radiated spectrum is overwhelmed by the  bright output of radioactive $^{56}$Ni  decay,  the signal in the radio, X- and gamma-ray  domains  may be detectable  from a few days to years, and - in Galactic SN remnants \citep[e.g.][see also Sect.~\ref{sec:SNRs_PWNe}]{Nayana2017} - up to  hundreds of years after explosion.}   

{Best candidates for  detection of radio and high-energy radiation are the core-collapse SNe whose progenitors have retained their hydrogen and helium envelopes before explosion (a.k.a. Type II SNe), which makes interaction with the shock and diffusive acceleration more effective. About 30\% of CC-SNe are detected at radio wavelengths, with the majority being Type II \citep[see review by][]{Kundu2026}. A fraction of these are also {X-ray-emitters, detected up to $\sim$100~keV \citep{Dwarkadas2025}. So far only one SN has been detected at MeV-GeV energies by Fermi-LAT \citep{Li:2026} from the nearby superluminous SN\,2017egm, which in fact may be detectable with CTAO \citep{Acero:2026}. Additionally, stringent upper limits have been determined for the bright nearby Type II SN2023ixf \citep{MartiDevesa2024,Ravensburg2024}.}}

{Similarly, only upper limits have been set so far to the TeV emission of core-collapse SNe by state-of-the-art Cherenkov telescopes \citep{2019A&A...626A..57H}.  The incoming CTAO may change this state of affairs thanks to its substantially improved sensitivity.  
It is expected that, while GeV emission from shock interaction with circumstellar medium may be detected by gamma-ray satellites during the first 2 weeks after explosion,  the simultaneous TeV emission may be suppressed by pair production owing to the initial large radiation density, and may be more likely detected later.}

{For comparable flux levels, CTAO is more sensitive than Fermi-LAT and will foreseeably detect with highest probability nearby (within 10 Mpc) Type IIP SNe, the most frequent SNe, and Type IIn SNe, that can be detected in a larger volume ($<$100~Mpc) owing to their higher brightness.} 

{It will then be possible to organize quasi-simultaneous campaigns with SKAO that will return accurate estimates of  circumstellar medium densities and mass-loss rates and in turn a detailed description of shock acceleration, estimate of potential cosmic-ray flux,  and a reconstruction of the  pre-explosion progenitor evolution.}

{Radio light curves allow estimates of the circumstellar density, mass-loss rates and shock velocity \citep{Corsi2017,Ho2020}, which in turn affect the maximum particle energies and very high energy photon fluxes. SKAO will enable high-sensitivity, high-cadence radio monitoring of nearby SNe, down to sub-mJy flux levels \citep{PerezTorres2015,Chandra01.2026.SKA,TaoAn02.2026.SKA}.  This will allow systematic searches for relativistic ejecta, constrain the circumstellar environments, and establish correlations with very high energy photon emission.} 

{A notable SN~IIP target of both SKAO and CTAO will be  SN1987A, with a predicted flux in 2027 of $\sim$4~Jy in the 150~MHz band. Even with a coarse beam of a $\sim$2 km core with perhaps only a dozen stations,
the SN will still stand out over the local LMC background.}

{Stripped-envelope, or Type Ib/c and Type IIb,  SNe are less frequent than Type II SNe,  but can release remarkable amounts of kinetic energy.  In rare cases ($\sim$5\%, almost all consisting in Type Ic SNe)  this is more than one order of magnitude in excess of the typical $10^{51}$ erg energy of SN explosion, which makes  these energetic SNe plausible candidates as  ultra-high-energy cosmic ray (UHECR) accelerators \citep{Chakraborti2011NatCo}.}  

{A subset of Type Ic SNe host GRBs \citep{Galama1998}, i.e. ultra-relativistic and highly collimated jets, observed within small ($<$5 deg) viewing angles. These are occasionally TeV emitters and, when located at redshifts less than $\sim$1 are detected by the Cherenkov telescopes  (see Sect.~\ref{sec:GRBs}).}   

{Extensive radio monitoring campaigns of nearby Type Ic SNe  have often revealed mildly relativistic ejecta that may be the signature of an inner engine \citep{Soderberg2010,Ofek2013,Corsi2017,Palliyaguru2019,Li2024,Chastain2025}.}
{This may have been responsible also for a mildly relativistic outflow although, despite their relative closeness ($<$50 Mpc),  no gamma-ray signal was detected by a high-energy satellite if one was present, see e.g. SN2009bb and SN2012ap  \citep{Margutti2014a,Margutti2014b,Chakraborti2015}.}   

{It is not clear whether in these cases a GRB was also produced, which was directed away from the line of sight and therefore not detected by the satellites, or a  jet was launched, but intrinsically weak or choked by the material immediately surrounding the explosion, so that  its energy was not sufficient to power an observable  GRB.  This suggests a continuum of explosion geometries and viewing angles.}

{The superior CTAO sensitivity  may afford detection of GRBs directed slightly off-axis \citep{Granot_Loeb2003} and exhibiting, being Doppler-disfavoured, lower TeV flux levels.  In these cases, the simultaneous radio and CTAO monitoring will map the evolution of the relativistic ejecta of the parent SN and the time-dependent activity of its inner engine.  This would represent an unprecedented result, as a unified scenario for GRBs associated with stellar core collapse has not received yet observational confirmation.}

{Our plan is to select very nearby (<50 Mpc) Type Ic radio-emitting SNe (occurring at a rate of no more than 1 per year) as targets of a high-cadence, accurate  monitoring with SKAO and CTAO with an aim to detect evidence of faint relativistic ejecta that can be compatible with a powerful, engine-driven explosion.}

    \subsection{X-ray binaries}




X-ray binaries (XRBs) are binary star systems in which a compact object, either a neutron star or a black hole, interacts with a stellar companion. This interaction can take place either through accretion of matter onto the compact object or in the interaction of the stellar wind with the relativistic outflows produced by the compact object. XRBs are among the most luminous X-ray sources in the sky and represent unique laboratories to study particle acceleration mechanisms and the impact of compact objects on their environments \citep{Harding1990, Fender2006, Rieger2007, Bykov2012, Paredes2013}.


A remarkable subset of XRBs hosting young non-accreting pulsars has been detected at gamma-rays by both space-born (e.g. \textit{Fermi}-LAT) and ground-based facilities (H.E.S.S., MAGIC, VERITAS, LHAASO, HAWC). The pulsar relativistic wind collides with the wind from the companion star, forming powerful shocks that accelerate particles up to multi-TeV energies. These gamma-ray binaries stand as some of the most efficient particle accelerators in our Galaxy, and provide critical insights into pulsar wind physics, stellar wind clumping, and radiation processes in extreme conditions. Prominent examples of these systems are PSR B1259--63 \citep{Aharonian2005}, LS 5039 \citep{Aharonian2006}, and LS I +61 303 \citep{Albert2006}. CTAO, with its improved sensitivity, angular resolution, and wider gamma-ray energy coverage, is expected to significantly enlarge this population \citep{Dubus2013}. Since gamma-ray binaries are long known to be also powerful radio-emitting systems (see e.g. \citealp{Paredes1990, Ribo1999, Gregory2002}), the combined capabilities of SKAO and CTAO will probe fainter systems, explore new classes of compact binaries, and deliver detailed phase-resolved spectra and light curves. Moreover, SKAO’s ability to detect and characterize the pulsar component in these systems will be crucial to interpret CTAO observations (see e.g. \citealp{Weng2022}). 
This will allow testing models of particle acceleration, discriminate between leptonic and hadronic emission scenarios, and understand the impact of binary parameters such as orbital eccentricity and stellar wind properties.

X-ray binaries displaying relativistic jets, dubbed microquasars, display SEDs extending from radio up to the VHE gamma-rays \citep{Bosch-Ramon2006}. Recently, they have also been proposed as PeVatron sources \citep{Abeysekara2018, Lhaaso2024,LHAASO2024b,Alfaro2024}. This UHE emission might be produced in some cases by particles reaccelerated when the jet impacts the surrounding medium \citep{Bosch-Ramon2005, Bordas2009}. Large-scale radio emission at the termination shock of microquasar jets is also expected (see e.g. \citealp{Heinz2002}), and has been detected in several systems \citep{Mirabel1992, Dubner1998, Marti2005, Pakull2010, Coriat2019, Motta2025, Atri2025}. The high-sensitivity of the next generation of Cherenkov telescopes together with the high-resolution, time-resolved radio imaging and polarimetry capabilities of SKAO \citep{Beri01.2026.SKA}, will bring new insights on the non-thermal outflows from these systems, both at binary-system length-scales as well as at the jet/medium interaction regions. 

Periodic and transient emission is characteristic from a variety of high-energy emitting High-Mass and Low-Mass XRBs (Be/XRBs, Supergiant XRBs, accreting milli-second PSRs, Colliding wind binaries, magnetars, or novae systems - see dedicated Sections in this Chapter), with durations that range from sub-second time-scales up to several years. Most of these systems remain however still undetected at VHE/UHE gamma-rays. The joint radio and gamma-ray wide-field coverage and flux sensitivity provided by CTAO and SKAO will open a new window to discover new transient/periodic XRBs, constraining some yet unknown basic properties, e.g. the emitter location or the acceleration mechanisms through which either leptonic, hadronic or combined particle populations radiate. The synergies between SKAO and CTAO data will thus provide a comprehensive view of XRBs, connecting the physics of compact objects, particle acceleration, and radiative processes from radio to very-high-energy gamma rays.



    \subsection{Novae}

        Novae are thermonuclear runaway explosions taking place at the surface of white dwarfs 
        when enough material is accreted from the companion star onto the compact object. The binary system survives the event, making it possible to restart the cycle of accretion and eruption \citep{BodeEvans_2012}.
        
        The energy release of the thermonuclear runaway explosion triggers an ejection of material and sustained nuclear burning, producing an enhanced emission across the electromagnetic spectrum from radio to gamma rays that lasts from days to years \citep{Chomiuk_2021}. Novae are excellent laboratories to test acceleration and radiative mechanisms because they are fast-evolving systems with an expected rate of 20--50 yr$^{-1}$ in the Milky Way \citep{De_2021,Kawash_2022,Zuckerman_2023}. Novae occurring in binary systems with main-sequence stars are called ``classical novae''. Conversely, if the companion star is a red giant star, these systems undergo ``symbiotic novae''. The term ``recurrent novae'' is employed for systems with more than one detected eruption. 
         
        In the radio band, novae present both thermal and non-thermal emission. The former is produced by the ionized expanding gas shell. This thermal Bremsstrahlung emission has characteristic timescales of years \citep[e.g.][]{Hjellming_1979}. The radio peak occurs after the optical maximum because the system remains optically thick at radio wavelengths. At early times, brightness temperatures exceeding 10$^4$~K provide evidence for shock-driven non-thermal synchrotron emission \citep[e.g.][]{Taylor_1987,Giroletti_2020}. In these shocks, particles are accelerated, reaching energies up to GeV--TeV, producing gamma rays that can reach TeV energies \citep{Ackermann_2014,HESS_2022,Acciari_2022,Abe_2025}. 
        
        Contemporaneous radio and gamma-ray emission
        has been detected in systems where the optical depth of the cool gas is relatively low compared to the values expected in typical classical novae \citep[e.g.][]{deRuiter_2023,Nyamai_2023}. Rapid follow-up of novae and monitoring of their emission is essential \citep{Lico01.2026.SKA}. A multi-wavelength picture of the non-thermal emission can shed light on the physical processes and radiative mechanisms in these rapidly evolving systems.  
        
        The radio band provides critical information on the temperature, morphology, and evolution of the ejecta. Monitoring at radio wavelengths allows us to probe the ejecta density profile and the total mass of the ionized medium. 
        Deep observations are required at early times because of the high free-free abortion with the ejecta \citep{Lico01.2026.SKA}. High-resolution images with radio interferometers at early times can provide insights into the origin of shocks and the mass ejection process \citep{Giroletti_2020,Lico_2024}. Such observations can deepen our understanding of the population of radiating particles and the physics of the shocks present in these systems. Moreover, high-resolution images can resolve the ejecta morphology, deriving more accurate estimates of its total mass. This quantity is important for gamma-ray emission, as systems with larger mass-loss rates are expected to present enhanced gamma-ray luminosity in the scenario of internal shocks.      
        
        The improved sensitivity and angular resolution of SKAO, together with improved sensitivity of CTAO at gamma rays, can help resolve fundamental questions about nova evolution, relevant radiative processes, and test acceleration mechanisms. In turn, particle acceleration and emission mechanisms in novae explosions are thought to occur also in the shocks of supernova explosions and in the early phases of supernova remnants (see dedicated Sections in this Chapter), providing complementary insights into the study of these latter sources.



    \subsection{Tidal Disruption Events}   

A tidal disruption event (TDE) is a rare transient that occurs when a star wanders too close to a massive black hole and is torn apart by tidal forces \citep{Rees1988}. TDEs produce a bright flare of electromagnetic radiation as the stellar debris is accreted by the black hole, illuminating a previously quiescent black hole. 
With the recent activation of all-sky optical and X-ray surveys, at the time of writing this chapter, there are $\sim$200 known TDE candidates, the majority of which have been discovered since 2015 \citep{vanVelzen2021,Yao2023,Hammerstein2023,Grotova2025}. Whilst TDEs are most commonly observed around SMBHs in the nuclei of galaxies, there are now a small number of candidate TDEs around off-nuclear massive black holes\citep[(MBHS; e.g.][]{Yao2025}, probing previously unobservable populations of black holes. 

At radio frequencies, TDEs provide a rare opportunity to witness the birth of jets and outflows around previously dormant MBHs as material is ejected either from shocks during the debris circularisation or from the subsequent accretion of material \citep[see][for a review]{Alexander2020}. Radio emission from TDEs arises from synchrotron emission produced by the shock between the ejected material and the environment directly around the black hole, or in some cases could arise from internal shocks within a relativistic jet. Radio observations of TDEs to date have revealed a dichotomy in the population of events, with a very small subset ($\sim1\%$) showing radio emission that is consistent with highly relativistic, collimated jets \citep[e.g.][]{Levan2011,Burrows2011,Zauderer2011,Bloom2011,Andreoni2022,Pasham2023}, while the majority of the population show significantly less relativistic ($\sim0.1\,c$) outflows that could be launched by debris collisions and shocks, super-Eddington accretion winds, or an unknown mechanism \citep[e.g.][]{vanVelzen2016,Alexander2016, Cendes2021,Goodwin2022, Goodwin2023a,Goodwin2023b,Goodwin2024,Christy2024}. 

The observed synchrotron-emitting outflows in TDEs provide direct evidence that particles are being accelerated to relativistic speeds, and demonstrate that TDEs provide the ideal site for the production of high energy gamma-ray emission. The relativistic electron population producing the synchrotron radiation in TDEs should also produce emission via inverse-Compton scattering, which should be detectable at gamma-ray energies \citep{Peng2016}. Tentative associations of high-energy neutrinos detected by IceCube with a couple of non-relativistic TDEs \citep{Stein2021,vanvelzen2024} suggest that TDEs provide the environment for PeV neutrino production, a necessary by-product of which would also be high energy gamma-ray emission. However, to date, there has been no detection of a TDE at GeV-TeV energies \citep{Peng2016}. Even TDEs in which clear multiwavelength evidence for an on-axis relativistic jet is present, no GeV emission was detected by Fermi \citep[e.g.][]{Burrows2011}. The GeV upper limits for the relativistic jetted TDEs constrains the gamma-ray emission from these TDEs to be less than $1\%$ that of their X-ray emission \citep{Peng2016}, which is a significantly lower flux ratio than found in gamma-ray bursts and blazars. 

While theoretical models show that GeV to TeV energy gamma-ray emission can be produced by both jet and non-jet mechanisms in TDEs \citep{Murase2020}, the lack of detected multi-GeV emission to date may be due to the absorption of the higher energy photons by an optically thick photosphere created by the stellar debris from the stellar disruption \citep{Wang2016}. Leading TDE models include a large optical/UV-bright photosphere in which much of the high-energy emission from the accretion disk is reprocessed \citep[e.g.][]{vanVelzen2016}. The presence of these lower energy photons imply a high optical depth of the source, preventing the gamma-rays from escaping \citep{Peng2016}.  

The SKAO will revolutionise our understanding of radio outflows from TDEs, with unprecedented sky-coverage and sensitivity to increase the number of known radio-emitting TDEs by an order of magnitude \citep{TaoAn03.2026.SKA,Shu01.2026.SKA}.
 CTAO constraints on any high energy emission from TDEs will provide crucial insight into the physical processes powering the radio outflows in these events, as well as constrain the presence of an optically thick photosphere in the case of further non-detections. Faint gamma-ray detections or deeper upper-limits of TDEs by CTAO will help constrain the optical depth and shock micro-physics that powers the synchrotron emission observed at radio frequencies, leading to a deeper understanding of the mechanism that powers outflows from accretion events on MBHs.

    \subsection{Monitoring of Blazars}    


It is now believed that most normal galaxies contain a supermassive black hole (SMBH) in their centre with a mass exceeding 10$^6$\,M$_\odot$. Active Galactic Nuclei (AGN) represent the evolutionary phase of evolution of SMBHs when they are actively growing via the accretion of circumnuclear material.  The gravitational energy released from this infalling matter can be converted into radiation, the kinetic energy of an outflowing plasma, or both. These outflows can form of narrow, well-collimated jets of plasma moving with relativistic speeds and emitting radiation observable from radio to gamma-ray energies and on scales from astronomical units to Mpc.

\begin{wrapfigure}{r}{0.7\textwidth}
    \centering

    \includegraphics[trim=50 20 50 20, width=0.7\textwidth]{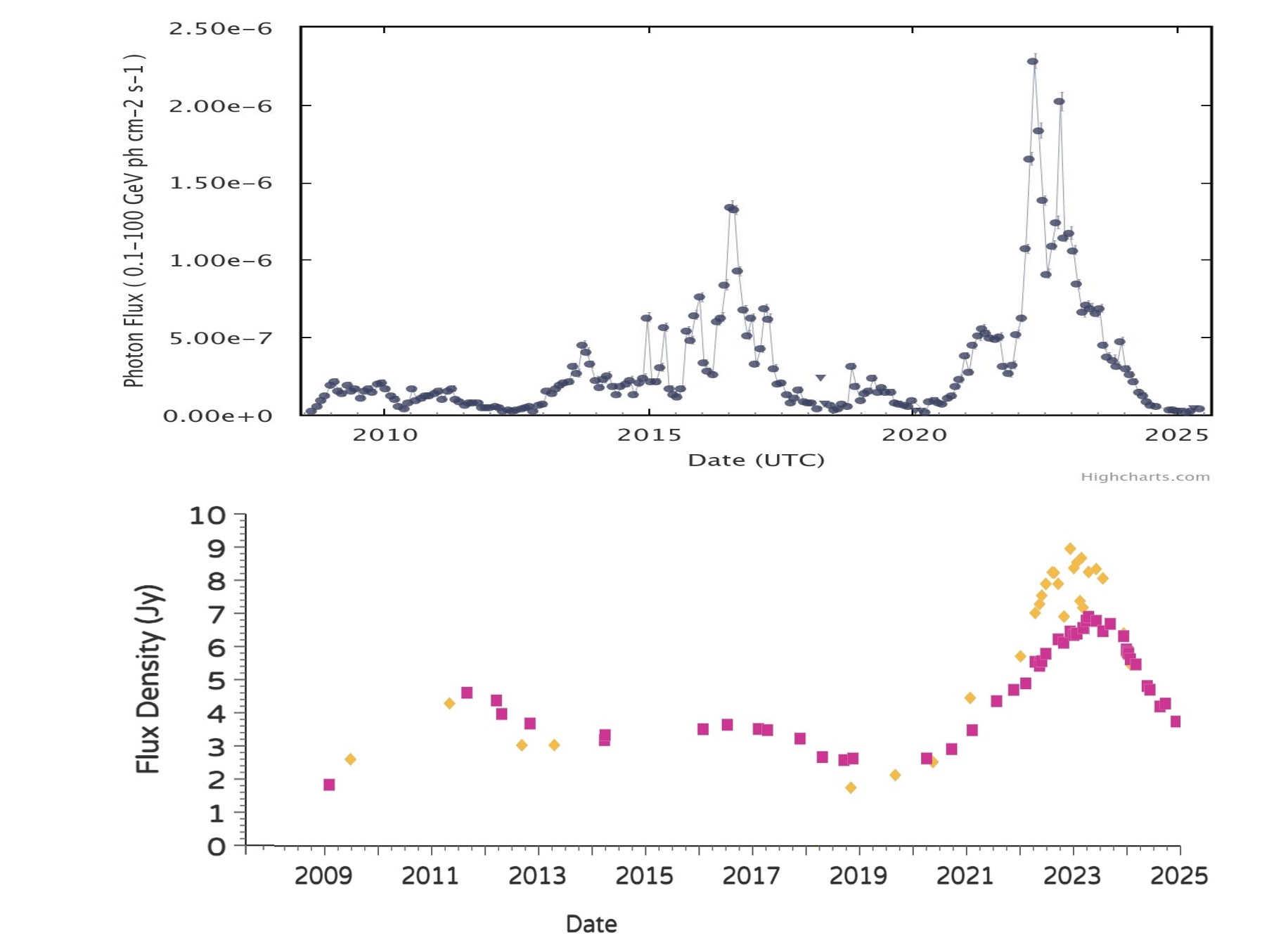}
    \caption{Fermi gamma-ray and ATCA radio light curves for the blazar PKS 0106+013 (4C +01.02). The top figure shows the Fermi light curve in one-month bins from the Fermi LAT Light Curve Repository \citep{Abd2023}. The bottom figure shows ATCA monitoring at 5.5\,GHz (purple) and 17\,GHz (gold), where statistical errors are smaller than the symbol size, and systematic errors are $\sim$5\% at 5.5\,GHz and $\sim$10\% at 17\,GHz. The under-sampled radio flare in 2009--2011 does
    not appear to follow any pronounced activity at gamma-ray energies, and the dramatic gamma-ray outbursts in 2015--2017 were only followed by a modest increase at radio wavelengths.
    However, the most recent gamma-ray flaring has seen the source flare with an increasingly inverted radio spectrum to record high flux densities.}
    \label{fig:PKS0106}
\end{wrapfigure}

Within the unified scheme of radio-loud AGN, blazars are identified as radio galaxies with powerful jets oriented at small angles ($<$ 5 degrees) relative to our line of sight. This unique orientation results in a variety of special relativistic effects within their ultra-fast plasma jets. 
 Blazars are characterized by variability on timescales from minutes to years, high degrees of radio, optical, and X-ray polarization, and commonly display apparent superluminal motion on the parsec-scale when monitored with VLBI. As an example, Figure~\ref{fig:PKS0106} displays the variable radio and gamma-ray light curves of the distant ($z=2.11$) blazar  PKS~0106+013 (i.e. 4C~+01.02).  The linear polarization associated with the non-thermal emission processes in blazar jets is notably high, often reaching 30\% - 40\% in optical and radio wavelengths \citep[e.g.][]{2006A&A...453..817V, 2010ApJ...710L.126M, 2012A&A...545A..48R}. These observed values represent a significant fraction of the theoretical maximum \citep{1979rpa..book.....R}. However, a disordered component within the magnetic field is necessary to explain the discrepancy between the theoretical maximum of $\sim$70\% (for uniform magnetic fields) and the typically observed values of around 10-15\% \citep[e.g.][]{2017ApJS..232....7F, 2023MNRAS.523.4504O} The optical polarization percentage of blazars exhibits high-amplitude variations, but these are not simply correlated with total light. This lack of simple correlation suggests that polarization may more directly and exclusively trace non-thermal radiation. Well-sampled variations in the polarization angle can provide valuable insight into the evolution and propagation of the plasma within these jets \citep{2008Natur.452..966M, 2010ApJ...710L.126M}.

 The complex phenomenology of blazar jets can be probed through dedicated multi-wavelength polarization monitoring programs using several facilities \citep[e.g.][]{2022Natur.611..677L, 2024A&A...689A.119K}. In particular, the linear and circular polarization parameters and their variability can be used to constrain the jet physical conditions, such as the magnetic field strength and topology, the particle density, and the plasma composition, and to study their dynamics. In this context, the monitoring of a sample of blazars with SKAO will be important for making a significant step forward to our knowledge.

Recent IXPE measurements of blazars indicate that their X-ray linear polarization can be high, variable, and not obviously correlated with observations at radio and optical wavelengths. This suggests that the emission region is not homogeneous and the electron population is energy-stratified. Specifically, in energy-stratified models, an electron population evolves over time as it propagates downstream from the acceleration region, resulting in a much higher X-ray polarization degree compared to optical and radio wavelengths. Therefore, contemporaneous monitoring of the source with radio, optical, and X-ray polarization measurements will be crucial to distinguish between the different emission mechanisms at play. The southern hemisphere combination of radio polarization measurements (SKA-mid having better sensitivity than SKA-low) with optical polarimetry on small- to mid-size telescopes (HIPPO in South Africa, the currently-prototyped CTApol in Australia, {and VSTPOL in Chile) will be able to deliver such a monitoring program, while X-ray satellites like IXPE and in the future new proposed missions, like e.g. the Enhanced X-ray Timing and Polarimetry (eXTP) and the Enhanced X-ray Polarimetry Observatory (EXPO), can provide polarization information in the X-ray band for a sample of blazars.} 
 
 Blazars constitute the majority of known extragalactic $\gamma$-ray emitters. Jets may be able to accelerate protons to EeV energies which contribute to the cosmic ray spectrum and be the origin of PeV neutrinos. For neutrino-emitting AGN, different levels of radio and optical polarization \citep{Zhang2019,Paraschos2025} are anticipated depending on whether a leptonic synchrotron or proton synchrotron scenario is at play, with the latter producing low values ($<$10\%). Operating at GHz frequencies, the SKA-mid is likely to be a key facility, with optical polarimetry, to understanding these multimessenger sources.

VLBI observations are important for understanding relativistic outflows from blazars. These observations offer direct evidence and are crucial for monitoring the motion of jet components on parsec scales \citep{Baczko01.2026.SKA}. By comparing parsec-scale variations with spectral energy distributions, polarization, Faraday rotation, multi-wavelength observations, and relativistic MHD simulations, we can determine key physical conditions within the jet and near the black hole. This includes jet composition, magnetic fields, variations in jet orientation, ages and densities of components, and the interplay between the jet and the ambient medium.
SKA will be instrumental in probing the connection between flares observed at VHE energies with CTAO and superluminal blobs and changes in magnetic fields over several years, as well as current-driven instabilities, tangled magnetic fields, or reconnection events. {In particular, SKA will provide unprecedented, high-sensitivity, high-resolution radio polarization maps, crucial for measuring the baseline jet magnetic field and Faraday rotation, identifying the site of the radio emission, and tracking potential long-term structural changes in the magnetic field to be compared with the activity observed at higher energies. An important connection between radio polarimetry with the SKA and optical/X-ray polarimetry for blazars lies in mapping the magnetic field structure and particle acceleration sites across different physical scales of the jet. While SKA probes the larger-scale, partially ordered magnetic field in the jet, optical and X-ray polarimetry can probe the intense, rapidly changing, and highly ordered magnetic fields near the acceleration site. 
This multi-wavelength polarimetric approach acts as a {\it magnetic tomography} of the jet, connecting the high-energy processes near the black hole to the lower-energy, extended emission regions.}

 
  





\section{Steady and Extended Sources}
\label{sec:steady_state_sources}
    \subsection{Planetary nebulae (PNe) and \textsc{H\,ii} regions 
    }

The basic model of \hii~regions and PNe, that of interstellar hydrogen gas photoionised by the UV flux from hot stars (young high-mass stars and old low-mass stellar cores respectively), does not obviously lend itself to gamma-ray observation. While the plasma is hot enough ($\sim10^4\,\mathrm{K}$) to emit flat-spectrum free-free (`thermal' or \emph{bremsstrahlung}) radio emission, the electrons do not have enough energy to generate gamma-rays \citep[e.g.\ ][see their Fig.~3]{2019A&A...630A..72P}. 
\hii~ regions are not generally associated with the strong magnetic fields that generate synchrotron emission in SNRs and accelerate cosmic rays.

However, gamma-ray emission is found towards these regions, and the basic models can be elaborated. \hii~ regions and PNe both arise from the combination of hot stars, that may be associated with cosmic-ray acceleration, with concentrations of interstellar gas that provides targets to generate gamma-rays under cosmic-ray bombardment. High UV fluxes also ionise the hydrogen and other atoms around them, which increases the gas pressure, leading to dynamic effects.

Although dominated by free-free emission, some \hii~ regions have been observed to have small amounts of synchrotron and gamma-ray emission, 
{which can be explained by shock acceleration of electrons in the \hii~region to $>$GeV energies \citep{2019A&A...630A..72P}. These high-energy electrons could generate gamma-rays, and, if the same mechanism were to accelerate protons, gamma-rays could also be emitted by pion decay. Leptonic gamma-ray production would be expected to be anticorrelated with radio synchrotron emission, and so the combination of resolution and sensitivity available to the SKA may allow for these mechanisms to be disentangled.}

Gamma-rays have been observed to arise from clusters of high-mass stars~\citep[e.g.~Westerlund\,1,][]{2022A&A...666A.124A}, through cosmic-ray acceleration either at the edge of the supershell or due to colliding stellar winds; or possibly due to the remnants of early supernovae. Sensitive high-resolution observations of the hot plasma in and around these clusters with SKAO could potentially map out the physical conditions and hence the gamma-ray emission mechanism, using not only the continuum emission but also radio recombination lines. \citet{2026JHEAp..5000465B} estimate the gamma-ray detectabilities of a few massive star clusters, assuming the gamma rays to arise from cosmic rays, themselves accelerated through the stellar winds. \citet{2016JHEAp..11....1H} used X-ray observations to look for a population of young stars in the \hii~ region G5.89--0.39, itself very close to W28 (mentioned above), so that the gamma-ray emission of the W28-cloud interface and of G5.89 may overlap.

White dwarf stars, similar to the stellar remnants found at the centre of PNe, can be gamma-ray sources due to thermonuclear detonation of accreted material (a nova) from a binary companion~\citep[e.g.~][and the relevant section of this work]{2025arXiv250107869D}. Nova eruptions could also generate gamma-rays by irradiation of the local interstellar medium (i.e.~the PN) --- 
high-sensitivity and high-resolution radio mapping of the local ionised gas possible with SKAO will allow the physical conditions to be mapped.

{These regions also emit spectral-line emission, both radio recombination lines and maser transitions.
Maser emission arises from several molecules; the ground-state transitions of the hydroxyl molecule (OH) are prominent, whereby the similar combination of collisional and radiative processes result in maser pumping and emission in both \hii~ regions and PNe alike. As a para-magnetic molecule, OH is also susceptible to Zeeman splitting under the influence of a magnetic field, and Zeeman splitting of OH maser emission can often readily identify field strength and direction for the in-situ magnetic field - there is thus the potential to tie the magnetic field information ascertained from the masers with the synchrotron emission observed at higher frequencies. OH Zeeman observations have been used to measure magnetic fields at the interface between the W28 SNR and nearby molecular clouds \citep{2005ApJ...620..257H}, a location with strong gamma-ray emission.}

Collectively this implies a scientific synergy for PNe and \hii~ studies between CTAO and, particularly, SKA-Mid, which can observe the free-free, synchrotron, OH maser (and associated magnetic fields) and radio recombination line emission with an unprecedented combination of resolution and sensitivity. Combining the SKAO with the resolution and sensitivity of CTAO offers the prospect of understanding not only the physical processes generating cosmic rays and gamma-rays, but also the effect of these high-energy messengers on the ISM around hot stars.


   \subsection{Supernova Remnants \& Pulsar Wind Nebulae}\label{sec:SNRs_PWNe}

\begin{figure*}[ht!]
\centering
    \includegraphics[trim=0 20 0 10,width=0.85\linewidth]{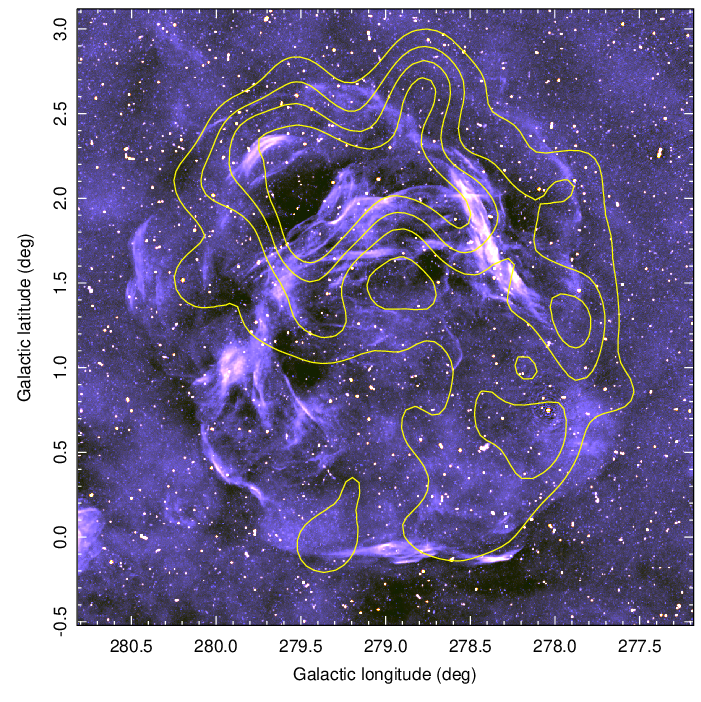}
    \caption{ASKAP 943\,MHz radio-continuum intensity image of Diprotodon (G278.94+1.35) -- one of the largest Galactic SNRs \citep{2024PASA...41..112F} seen at gamma-ray and radio frequencies. The synthesized beam for the radio image is $15^{\prime\prime}\times15^{\prime\prime}$. 
    The \textit{Fermi}-LAT contours, for energies above 1~GeV, correspond to 5, 10, 15 and 20~$\sigma$ significance levels (smoothness=6). 
    }
    \label{fig:3}
\end{figure*}

Supernova remnants (SNRs) are among the most prominent non-thermal radio sources, with over 300 currently catalogued~\citep{2025JApA...46...14G, 2012AdSpR..49.1313F}. Current radio surveys with the SKAO precursors have already demonstrated remarkable potential to discover and analyse SNRs, particularly in the field of discovering low angular size and surface-brightness SNRs which were previously missed~\citep{2023MNRAS.524.1396B,2025ApJ...988...75B,2025A&A...693A.247A,2025PASA...42...21M,2022MNRAS.512..265F,2024PASA...41..112F,2024RNAAS...8..158S,2024MNRAS.534.2918S,2026arXiv260416882L}, some of which also display gamma-ray emission~\citep{2025PASA...42..104F,2023AJ....166..149F,2024A&A...684A.150B} such as most recent Diprotodon SNR \citep[G278.94+1.35; see Fig.~\ref{fig:3}, taken from][]{2024PASA...41..112F}. SKAO's improved sensitivity and resolution will push this field even further \citep{Ingallinera01.2026.SKA,Gelfand01.2026.SKA}, helping to uncover the faintest and most evolved remnants, and potentially also extremely low angular-size young supernova remnants and pulsar-wind nebulae \citep[PWNe,][]{2024PASA...41...32L}. SKAO also has the potential to resolve the brightest members of SNR populations in even further galaxies than currently possible, for example our nearby neighbours the Magellanic Clouds (MCs) are prime targets with well-studied SNR populations~\citep{2024MNRAS.529.2443C,2021MNRAS.506.3540P,2017ApJS..230....2B,2023MNRAS.518.2574B,2024A&A...692A.237Z}. This will provide more complete samples for population studies and investigating the origin of cosmic ray production; an area of active investigation which requires collaboration with the next generation of high-energy telescopes. 

The synergy between radio and high-energy observations is one of the most powerful, and is crucial for advancing our understanding of cosmic ray and gamma ray production in the Universe~\citep{2021map..book.....F}. One of the key challenges with high-energy observations is differentiating between leptonic and hadronic emission. Radio-continuum observations can help distinguish between. Whole SED modelling, using data from radio to gamma-rays, has been demonstrated as an effective tool for differentiating between hadronic and leptonic processes~\citep{2010ApJ...712..287E,2011ApJ...734...28A,2012A&A...538A..81M}. 3D models of the particles escaping SNRs and interacting with surrounding gas  \citep[e.g.][]{Einecke-W28:2024,Rowell-RXJ1713:2024} will also be particularly important in this discrimination. Strong spatial correlation between radio (and X-ray) and gamma-ray emission can suggest that both arise from the same electron population through synchrotron and IC processes~\citep{2008ARA&A..46...89R}. Poor correlation may indicate either different electron populations, varying local conditions, or the presence of hadronic gamma-ray processes. Additionally, radio observations can help constrain the environment and magnetic field, which is crucial information for distinguishing between these emission mechanisms. This synergy has been used in the analysis of several SNRs which emit at both radio and gamma-ray frequencies~\citep{2024A&A...684A.150B,2020A&A...643A..28D,2024A&A...689A.257L,2023ApJ...954....1C}. The radio-continuum correlation is integral to uncovering information about an SNR's properties, such as age, environment, and potential interaction~\citep{2018MNRAS.475.5237G,2023ApJ...954....1C}. Additionally, PWNe can produce both radio, X-ray and gamma-ray emission through synchrotron and IC processes~\citep{2017hsn..book.2159S}. PWNe represent the most populous source type in the TeV gamma-ray band \cite{HESS-PWN:2018}. The improved sensitivity and resolution of both SKAO and CTAO can enable more detailed correlation studies to map electron energy distributions and trace the overall PWN structural evolution. CTAO is one of the upcoming telescopes that will provide an unprecedented view of the gamma-ray sky, allowing a better window into our own Galaxy and the nearby MC system~\citep{2023MNRAS.523.5353A}. Additionally, the upcoming SWGO will be complementary to SKAO observations. The SWGO will have an almost continuous 24\,hr operation and a wide field of view which allows it to observe the entire overhead sky~\citep{2024JInst..19C2065C}. This makes it an excellent survey tool when paired with SKAO's survey capabilities, particularly for analysing large angular-size SNR, such as Antlia or G354$-$33~\citep{2021ApJ...920...90F,2002ApJ...576L..41M}, which may require mosaicing observations with CTAO. SKAO's ability to detect and characterise large SNR populations, combined with systematic CTAO follow-up, can help resolve fundamental questions about SNR evolution, CR acceleration efficiency, and the influence of age and SNR type on leptonic to hadronic emission mechanisms.

    \subsection{Odd Radio Circles}

Odd Radio Circles (ORCs) are a recently discovered class of objects, so far exclusively seen at radio frequencies~\citep{2021PASA...38....3N, 2022MNRAS.513.1300N}. They are generally characterised by their circular or ring-like morphology and low radio surface brightness, and most often have elliptical galaxies located near their geometric centre~\citep[][]{2021PASA...38....3N, 2022MNRAS.513.1300N, 2021MNRAS.505L..11K, 2023MNRAS.520.1439L, 2023ApJ...945...74D}, with a few notable exceptions~\citep[][MacGregor et al. in prep]{2022MNRAS.513.1300N, 2024MNRAS.531.3357K}. There are also several ORC candidates discovered~\citep[][Filipovi\'c et al. in prep]{2022PASA...39...51G, 2025PASA...42...97G}, as well as some objects of other classes that can appear morphologically similar~\citep{2024MNRAS.532.3682K}, some of which were initially classified as candidates~\citep{2022MNRAS.512..265F}. The true origin of ORCs is still debated with several different scenarios proposed, including AGN activity such as precessing jets~\citep{2023ApJ...948...25N} or re-energised remnant lobes~\citep{2024PASA...41...24S}, galaxy merger shocks~\citep{2023ApJ...945...74D}, SNR~\citep{2022MNRAS.512..265F}, black hole mergers~\citep{2022MNRAS.513.1300N}, and tidal disruption events~\citep{2022MNRAS.516L..43O}. ORCs have been mainly discovered by the ASKAP, LOFAR, and MeerKAT telescopes, primarily due to their excellent surveying capabilities~\citep{2009ASPC..407..446J,2021PASA...38....9H, 2013A&A...556A...2V, 2016mks..confE...1J}. This surveying capability is particularly well highlighted by recent work, which has found ORC candidates in great numbers by using machine learning algorithms on the vast amount of survey data currently available~\citep{2022PASA...39...51G, 2025PASA...42...97G}. SKAO stands to improve this current capability substantially, with improved sensitivity, resolution, and survey speed, allowing ORCs and ORC candidates to be uncovered in greater numbers.

The most significant advantage of SKAO synergy with high-energy telescopes, such as the upcoming CTAO, will be in helping to constrain the ORC emission mechanisms. If a gamma-ray signature is detected, then determining whether the emission is hadronic, leptonic, or a mixture of both will be integral to uncovering the underlying physics and differentiating between the competing origin scenarios. Strong gamma-ray and radio-continuum correlation would favour leptonic scenarios, such as scenarios with active particle acceleration, while anti-correlations with radio will favour hadronic processes or scenarios potentially involving older electron populations. Higher sensitivity gamma-ray observations are necessary to distinguish between these emission mechanisms, but large-scale radio surveys, such as those conducted by SKAO, are needed to provide the targets and guide these more targeted observations \citep{SabyasachiPal02.2026.SKA,Koribalski01.2026.SKA}. Whether CTAO is able to detect ORC gamma-ray counterparts or not, even a negative detection can establish upper limits, helping to constrain the emission mechanism. The ability of SKAO to uncover a larger ORC population, combined with high-energy follow-up, will enable us to determine the origin of ORCs, and even whether all these objects represent a single astrophysical phenomenon or multiple distinct origins.

\subsection{Mapping Magnetic Fields in the Magellanic Clouds}

The Large and Small Magellanic Clouds (LMC and SMC) are among the closest galaxies to the Milky Way, at distances of approximately 49.5~kpc and 62.8~kpc, respectively. Their mutual gravitational interaction induces tidal forces that strip gas and stars from the SMC, forming the Magellanic Stream and Bridge \citep[][and references therein]{Bruns2005,Luri2021}. Notably, the Magellanic Stream is one of the largest observable structures, {consisting of a trail of high-velocity gas clouds extending more than 100$^\circ$ across the sky and trailing behind the LMC and SMC.} Altogether, the proximity of the Magellanic Clouds, as well as the Magellanic Stream and Bridge, {makes this system an ideal target} for both the SKAO and ground-based TeV facilities such as the CTAO and future SWGO.

The recent Gaia Data Release 3 has provided further insight into the LMC and SMC. \citet{Luri2021} traced the stellar component of the Magellanic Bridge in both spatial and kinematic dimensions, while also revealing additional substructures in the outer SMC. \citet{Ma2023} analysed the orientation of SMC H~I filaments from the ASKAP survey in relation to the magnetic field traced by optical polarization, finding a strong alignment. In an earlier study, \citet{lobo-gomes2015} mapped optical polarization across the SMC’s northeastern (NE) body, the Wing, and parts of the Bridge, revealing the underlying magnetic field morphology. {They identified several ordered components along the NE body and Bridge; however, the overall magnetic field structure remains complex and not yet fully understood.}

{Overall, current observations suggest that the magnetic field in the Magellanic system is structured on multiple scales but remains difficult to interpret due to the interplay between tidal interactions, gas dynamics, and star formation activity. Future instrumentation is expected to significantly} improve this picture. VSTPOL, a linear optical wide-field imaging polarimeter, will soon represent a major upgrade of the VLT Survey Telescope (VST) in Paranal, Chile \citep{Schipani2024}. Thanks to its wide field of 1~degree$^2$ and high spatial resolution of 0.21~arcsec, {it will enable detailed optical polarization maps of the Magellanic Clouds, providing new constraints on both the structure and strength of their magnetic fields} \citep{VSTPOLcoll2026}. Additionally, \citet{2024MNRAS.535.1944L} mapped the magnetic field of the LMC and found it to be distinct from the SMC, Magellanic Bridge, and larger system, {highlighting the complexity of the system and suggesting different evolutionary histories for its components.}

The SKAO is expected to provide unprecedented sensitivity and resolution in the radio domain. This will allow high-resolution mapping of the H~I filaments previously detected by ASKAP, while radio polarization observations may reveal the underlying magnetic field orientation relative to large-scale structures. {In particular, SKAO observations will help establish a direct connection between gas dynamics and magnetic field geometry across multiple spatial scales \citep{Ma01.2026.SKA,Mao01.2026.SKA}.} SKAO will likely resolve the complex morphology of the filaments, shaped by internal gas motions, tidal interactions with the LMC, and stellar feedback. These maps can be compared with future measurements of the SMC magnetic field from starlight and dust polarization, as well as diffuse polarized synchrotron emission. For example, synchrotron emission at 1.4~GHz is primarily produced by 5~GeV cosmic-ray leptons in a typical 4~$\mu$G interstellar magnetic field \citep{Blumenthal_Gould1970}, with the total radio flux measured from ATCA~+~Parkes observations being 426~Jy \citep{Hughes2007}, including $\sim$50~Jy from background sources and $\geq$20\% from thermal bremsstrahlung \citep{2023MNRAS.523.5353A}. {Combining radio observations with gamma-ray data will provide complementary constraints on star formation activity and interstellar medium properties.}

Indeed, in the context of high-energy astrophysics, the LMC is one of the few external star-forming galaxies that can be spatially resolved. At GeV and TeV energies, only the Magellanic Clouds and Andromeda provide sufficient angular resolution to enable detailed studies \citep{Abdo2010_LMC_Fermi,Abdo2010_SMC_Fermi,Ackermann2016,Ackermann2017,Acero2009,Abdalla2018}. Its large angular extent, low inclination, proximity, and active star formation make the LMC particularly favourable for such analyses. It hosts several remarkable astrophysical objects, including 30\,Doradus --- the most luminous H~II region in the Local Group; PSR\,J0537--6910 --- the most energetic known pulsar; and SN~1987A, the remnant of the nearest modern core-collapse supernova. All of these sources are confirmed or potential sites of particle acceleration and gamma-ray emission.

At high and very-high energies, the LMC has been characterized by Fermi-LAT and H.E.S.S., revealing five point sources, three of which are detected in both GeV and TeV ranges: the pulsar PSR\,J0537–6910 and its nebula, the supernova remnant N\,132D \citep{2020ApJ...902...53S}, and the gamma-ray binary LMC\,P3. The remaining two sources include PSR\,J0540–6919 \citep{2014ApJ...780...50B}, detected only in GeV gamma rays, and the superbubble 30\,Doradus~C \citep{2019A&A...621A.138K,2021ApJ...918...36Y}, detected only at TeV energies \citep{Ackermann2015,Ackermann2016,Abramowski2015}. The LMC also exhibits diffuse, galaxy-scale emission of likely interstellar origin, produced by the population of cosmic rays, with additional kpc-scale components of uncertain origin in regions apparently devoid of gas \citep{Ackermann2016}. These extended emissions are currently observed only in the $\sim$100~MeV–100~GeV range, leaving higher-energy behaviour largely unexplored. Emission between $\sim$100~GeV and 100~TeV probes more energetic cosmic rays and earlier stages of their life cycle, as high-energy particles escape more efficiently via diffusion.
Remarkably, one of the Key Science Projects (KSPs) of the CTAO Consortium involves a deep survey of the LMC \citep{2023MNRAS.523.5353A}. 
The scientific goals include population studies of various object classes, investigations of the interstellar medium and galactic cosmic rays, and indirect searches for dark matter. Concerning the last, the combination of CTAO monitoring of the LMC with deep SKAO observations can potentially improve existing constraints on the cross section for WIMP particles, so far obtained with SKAO precursors such as ASKAP \citep{Regis2021}.



    \subsection{Nearby Starbursts and Radio Galaxies}

Starbursts and radio galaxies are extra-galactic sources characterized by significant radio emission at low frequencies in radio. They typically have 1.4~GHz luminosity densities lower and higher than 10$^{30}$~W~Hz$^{-1}$, respectively. The dominant emission mechanisms responsible for such low frequency radio emission is non-thermal synchrotron emission originated by relativistic particles. This is thought to be mostly produced by electrons accelerated in supernova remnants of massive stars, in the case of starbursts, or relativistic particles emitting in the proximity of the AGN and then associated with the radio core and large scale jets. Interestingly, ongoing and forthcoming ground based gamma-ray facilities such as the CTAO may detect several classes of extragalactic sources in the TeV domain. Among them there are also powerful starburst galaxies and radio galaxies, which in nearby cases such as Centaurus\,A, M87, Cygnus\,A, and IC\,310 have demonstrated the ability to accelerate particles up to very high energies, ultimately producing TeV photons. Of particular interest is the type~II AGN NGC~1068, the only confirmed non-blazar source of TeV neutrinos so far. Its observed neutrino flux surpasses the expected TeV gamma-ray flux by more than an order of magnitude \citep{Icecube2022}.

The synergy between upcoming radio facilities such as the SKAO and ground-based TeV observatories like CTAO offers a powerful avenue to investigate the high-energy emission mechanisms in nearby starbursts and radio galaxies \citep{Bruni01.2026.SKA}. The unprecedented sensitivity and resolution at low radio frequencies of SKAO enable detailed mapping of synchrotron emission from cosmic-ray electrons and thermal free–free emission from H~II regions, providing insights into the distribution and transport of cosmic rays. At the same time, CTAO will probe the corresponding TeV photon output, which traces particle acceleration up to extreme energies. By combining the spectral and spatial information from SKAO with the TeV detections (or upper limits) provided by CTAO, it will be possible to disentangle leptonic and hadronic processes, assess the role of supernova remnants and active galactic nuclei in particle acceleration, and constrain the efficiency of cosmic-ray transport in dense galactic environments. This multi-wavelength approach will therefore be key to establishing a consistent picture of the origin of high-energy emission in star-forming and radio galaxies \citep[e.g.][]{Eichmann2016,Ohm2016,Boccardi2019,Rani2019,Rulten2022}.

    \subsection{Fermi Bubbles and the Galactic Centre}
    

The Fermi Bubbles are giant, lobe-like structures that emerge into north and south Galactic hemispheres from the inner Galaxy extending to $\|b\| \sim 55^\circ$.
They were first discovered \citep{Su2010} in gamma-ray data, but they have counterparts \citep{Sarkar2024} at other wavelengths, including in polarized, $\sim$GHz radio continuum emission \citep{Carretti2013}.

%
The SKAO will be a very powerful instrument at  scales $\lesssim$ degree (in  L-band), including probing structures in the  
Galactic centre \citep{Schoedel01.2026.SKA} such as the non-thermal filaments or bubble-like structures in other galaxies.   Assuming use of Band 2 of SKAO-MID, centred at 1.31 GHz with 0.72 GHz of bandwidth, and an exposure time of 10\,h, the sensitivity calculator returns a resolution of 0.2\,arcsec and a sensitivity of 4 $\mu$Jy/beam, which turns to 1$\mu$Jy/beam at a coarser resolution of 1\,arcsec. These values are significantly better than those obtained by MeerKAT in the Galactic centre, where the SKAO can see finer and fainter details. 

Given this sensitivity and resolution, 
an interesting target for studies of the Fermi Bubbles and the Milky Way's nuclear outflow in general would be the relatively dense, neutral (\hi~ and molecular hydrogen) gas clouds entrained into this outflow \citep{DiTeodoro2018}.
These clouds present target nuclei with which the hadronic (i.e, proton and heavier ion) cosmic ray population putatively \citep{Crocker2011,Crocker2015,Sarkar2024} suffusing the Bubbles (and/or entrained into the nuclear outflow and possibly even helping to accelerate it) may collide, ultimately producing both direct $\gamma$-rays via the proton-proton collision mechanism.
Of most relevance here are the high energy `secondary' electrons and positrons {as mentioned in Section\,\ref{Intro}. The resulting secondary synchrotron emission
%
can be emitted on the clouds' entrained magnetic field in $\sim$GHz radio continuum band. The following discussion looks at SKAO's potential to detect this emission from the Fermi Bubbles.}

%
For a cloud of mass $M$ permeated by CR protons of energy density $u_p$, the total $pp$ energy-loss power is
\begin{equation}
\dot{E}_{pp} \;=\; u_p\,\kappa\,\sigma_{pp}\,c\,\frac{M}{\mu m_p}\,\phi_{\rm pen}\,,
\end{equation}
with mean molecular weight $\mu\simeq1.4$, $\kappa \sim 1/2$ is the inelasticity of $pp$ collisions, and where $\phi_{\rm pen}\le1$
is the CR penetration fraction of cloud mass which accounts for partial exclusion/skin-depth effects (and represents a significant uncertainty of this analysis). The secondary electron/positron injection power is
\begin{equation}
\dot{E}_{e,\mathrm{inj}} \;=\; f_e\,\dot{E}_{pp}\,,
\end{equation}
where $f_e \sim 1/6$.
Electrons of Lorentz factor $\gamma$ synchrotron radiate near the characteristic frequency
\begin{equation}
\nu \;\simeq\; 4.2\,B_{\mu\mathrm{G}}\,\gamma^2~\mathrm{Hz}\,,
\end{equation}
so $\gamma\simeq5.8\times10^{3}$ (or $E_e \sim$ 3 GeV) at $\nu=1.4~\mathrm{GHz}$ for $B=10~\mu\mathrm{G}$. Let the synchrotron loss fraction at the relevant energy, $E_\nu$, be
\begin{equation}
f_{\rm syn}(\nu)\;=\;\frac{b_{\rm syn}(E_\nu)}{b_{\rm syn}+b_{\rm IC}+b_{\rm brem}+b_{\rm Coul}}\Bigg|_{E_\nu}\!,
\end{equation}
with $U_{\rm rad}\!\sim\!1~\mathrm{eV\,cm^{-3}}$ and gas density $n$ setting $b_{\rm brem},b_{\rm Coul}$. Representative values of $f_{\rm syn}$ at $1.4~\mathrm{GHz}$, for $n = 10$ cm$^{-3}$ and $n = 100$ cm$^{-3}$, are: $(B=10~\mu\mathrm{G}) \, f_{\rm syn}\approx0.083 \,\&\, 0.009;  (B=30~\mu\mathrm{G} ) \, f_{\rm syn}\approx0.32 \,\&\, 0.046; (B=100~\mu\mathrm{G}) \, f_{\rm syn}\approx0.74 \,\&\, 0.23$


For a steady-state power-law $e^\pm$ spectrum with index $p\approx2.2$, the synchrotron power per logarithmic frequency interval near $\nu$ is a fraction $\xi_\nu\sim0.1$ of the total synchrotron power. Normalising to fiducial values, the resulting specific flux density at Earth is
then \begin{eqnarray}
S_\nu &\simeq&  0.15~\mathrm{Jy}\; \label{eq:Scalebox} \\
&\times&
\Big(\frac{u_p}{1~\mathrm{eV\,cm^{-3}}}\Big)
\Big(\frac{M}{10^5\,M_\odot}\Big)
\Big(\frac{\phi_{\rm pen}}{1}\Big)
\Big(\frac{f_{\rm syn}}{0.01}\Big)
\Big(\frac{\xi_\nu}{0.1}\Big)
\Big(\frac{1.4~\mathrm{GHz}}{\nu}\Big)
\Big(\frac{8.2~\mathrm{kpc}}{D}\Big)^2
\Big(\frac{\sigma_{pp}}{50~\mathrm{mb}}\Big)
\Big(\frac{\kappa}{0.5}\Big)
\Big(\frac{f_e}{1/6}\Big) \nonumber
\,.
\end{eqnarray}

Overall, we find with this relation that, for  plausible or even conservative GC-wind parameters $u_p\sim10^{-2}\!-\!10^{-1}~\mathrm{eV\,cm^{-3}}$, clouds with $M\gtrsim10^5\,M_\odot$ are detectable with SKA1-Mid at GHz frequencies if $B\gtrsim10\!-\!30~\mu\mathrm{G}$ and $\phi_{\rm pen}\gtrsim0.03$.
%
For instance, taking $D=8.2~\mathrm{kpc}$, $U_{\rm rad}=1~\mathrm{eV\,cm^{-3}}$,
a compact, massive, but modestly magnetised cloud with $M=10^5\,M_\odot$, $R=5~\mathrm{pc}$ (diameter $\simeq2.1'$), $n\simeq100~\mathrm{cm^{-3}}$, 
and
$B=30~\mu\mathrm{G}$ 
gives
$f_{\rm syn}\approx0.046$.
Using Eq.~\ref{eq:Scalebox}, for
$u_p=0.1~\mathrm{eV\,cm^{-3}}$ and $\phi_{\rm pen}=0.1$ this implies
$S_{1.4}\approx6.9~\mathrm{mJy}$. Distributed over $\sim4.4\times10^2$ beams at $10^{\prime\prime}$, the mean surface brightness $\sim16~\mu\mathrm{Jy\,beam^{-1}}$
which should mean a
$\gtrsim8\sigma$ detection in 1~h.
Radio continuum observations of the known clouds in the Fermi Bubbles/nuclear wind with the SKAO thus hold out the interesting prospect of constraining the cosmic-ray content of those structures.

     \subsection{Galaxy Clusters}

Radio observations of galaxy clusters have revealed steep-spectrum diffuse synchrotron sources, extending even beyond the Mpc-scale, which indicate the presence of relativistic electrons and magnetic fields mixed with the intracluster medium (ICM) \citep{vanweeren19rev}. The origin and evolution of these non-thermal components are still under investigation, making their study an active field of astrophysical research \citep{brunetti14rev}. In recent years, SKAO pathfinders and precursors have significantly expanded the census of diffuse radio sources in clusters and opened new parameter spaces to explore their properties \citep{govoni19, digennaro21fast, botteon22dr2, knowles22, duchesne24psz2, phuravhathu25}. With the advent of SKA-LOW and SKA-Mid, this field is expected to flourish even further, while the synergy with the next-generation of $\gamma$-ray instruments will be crucial to understand the origin of cluster radio sources \citep{Cassano01.2026.SKA,Gitti01.2026.SKA,Vacca01.2026.SKA,ArpanPal01.2026.SKA}. {Indeed, current claims of gamma-ray emission from clusters are disputed, and the improved sensitive of CTA could provide the first clear detections \citep{cta_book2019,Abe_2024_Perseus}.} 

Diffuse sources in clusters are broadly classified as radio relics, halos, and mini halos:

\begin{itemize}[leftmargin=1em]
    \item \textbf{Radio relics} are generally observed in the outskirts of galaxy clusters, show significant linear polarization, and usually have an elongated, arc-like, morphology characterized by sharp emission edges. They are believed to be tracers of shock fronts driven during cluster mergers, where cosmic ray electrons are (re)accelerated. As shocks in galaxy clusters have low Mach numbers ($\sim$2--3), this poses a problem on the efficiency of the acceleration process and on the electron-to-proton ratio injected at shocks \citep{botteon20efficiency}. In this case important constraints on the acceleration process and on the energy partition between protons and electrons can be obtained by combining radio and $\gamma$-ray observations (limits) \citep{vazza14challenge, wittor21rev}.

    \item \textbf{Radio halos} are observed at the centre of galaxy clusters. Their emission is unpolarized down to a few percent level, broadly follows the distribution of the thermal gas, and permeates the volume of the ICM. Current observations favour the scenario for which CRe in halos are re-accelerated by turbulence in the ICM driven by merger events \citep{brunetti01coma}. The alternative scenario, the hadronic model \citep{dennison80, blasi99}, in which nuclear collisions between CRp and the nuclei of the thermal ICM continuously produce secondary electrons, is disfavoured by \textit{Fermi} upper limits to the $\gamma$-ray emission from galaxy clusters, which put stringent constraints on the role of hadron-hadron collisions and thus on the $\pi^0$ generation \citep{ackermann14, brunetti17, Adam2021, Osinga2024}. As the direct acceleration of electrons from the thermal ICM pool is a very inefficient mechanism, a population of seeds CRe distributed on the cluster-scale is required. Secondary electrons generated by hadronic collisions may constitute a significant fraction of the seed CRe; in this case, the $\gamma$-ray signal is predicted at a lever that is just below current upper-limits \citep{brunetti17, Adam2021}. {Furthermore, recent observations with MeerKAT, the main precursor of SKA-Mid, are unveiling the presence of cluster-scale radio halos in relaxed galaxy clusters which in principle would be consistent with a hadronic scenario \citep{vanWeeren2026} and improving gamma-ray observations will be crucial to determine its origin.}

    \item \textbf{Mini halos} are typically confined within the central cooling region of relaxed clusters. Similar to halos, turbulent re-acceleration and hadronic models have been proposed to explain the origin of mini halos \citep{gitti02, pfrommer04hadronic}. As clusters hosting mini halos do not show significant levels of dynamical disturbance, re-acceleration models require more gentle processes than merger activity to inject turbulence in the ICM in order to preserve the cool-core region. These motions may be connected with the inward advection of turbulent matter within the cooling flow region or due to the sloshing generated as a consequence of minor/off-axis mergers \citep{zuhone13}. Secondary models are also particularly attractive in mini halos as their radio emission traces the denser regions, namely where the thermal targets number density is larger \citep{pfrommer04hadronic, zuhone15}. The primary protons for the hadronic collisions could be naturally provided by the powerful central AGN often residing at the cluster cluster. The discrimination between re-acceleration and hadronic models for radio mini halos formation is still very challenging. Current $\gamma$-ray upper limits are not yet able to put significant constraints on the origin of these radio sources, being still consistent with a purely hadronic origin of mini halos \citep{perkins06, aleksic12, ignesti20connection}. Nonetheless, the connection between mini halos and sloshing features such as cold fronts \citep{mazzotta08} and the existence of correlations between the mini halo and the central AGN properties have also been observed \citep{richardlaferriere20}.
\end{itemize}
   
    \subsection{Dark Matter and Beyond-Standard-Model Physics}

A key mission for gamma-ray and radio facilities is searching for particle dark matter signatures~\citep{Regis01.2026.SKA, CTA:2020qlo,CTAO:2024wvb,CTAO:2025gdd,Colafrancesco:2015ola}. If dark matter consists of Weakly Interacting Massive Particles (WIMPs), it can annihilate or decay into Standard Model particles, producing photons through several channels. Direct, loop-mediated production creates an unambiguous spectral line at the dark matter mass, while annihilation into quarks or gauge bosons yields a photon continuum bounded by the WIMP mass \citep[see][for a recent review]{Balazs:2024uyj}. Additionally, charged leptons produced in these processes generate lower-energy photons via synchrotron radiation, inverse Compton scattering, and bremsstrahlung.

For GeV-TeV mass WIMPs, these processes yield a mix of photons from gamma-rays to radio waves, creating a continuous flux from dark matter-rich regions like the Galactic Centre (GC) and dwarf galaxies. Combining radio and gamma-ray observations is crucial, especially for the GC, due to significant astrophysical backgrounds. These include diffuse emission from cosmic rays, known sources, the Fermi Bubbles, and unresolved sources like a hypothetical millisecond pulsar population that can mimic WIMP signals. Such backgrounds are subject to major astrophysical and ontological uncertainties.

Crucially, any background model gives concrete radio and gamma-ray predictions for spectral energy distribution and morphology. While WIMP annihilation signals share some uncertainties with backgrounds (plus others, like the dark matter density distribution), their multi-wavelength signatures differ markedly from astrophysical predictions, enabling discrimination even when a single band appears similar. The reasons for combining radio and gamma-ray studies are thus compelling.

\begin{itemize}[leftmargin=1em] 
\item Combining data improves sensitivity by better discriminating WIMP signals from astrophysical backgrounds. A discovery would provide crucial information for determining the correct dark matter model, as the relative proportion of annihilation products is unconstrained \emph{a priori} \citep{Crocker:2010gy, Balazs:2014jla}.
\item Where WIMP signals are sub-dominant, multi-wavelength data can break degeneracies in background models, for example, using radio to constrain magnetic fields while examining other components in gamma-rays \citep{vanEldik:2015qla}.
\item Pure astrophysical studies, like tracing gas in radio or examining potential millisecond pulsars in gamma-rays, can be used to seed refined background models for a combined analysis of all data \citep{Crocker:2010gy}.
\end{itemize}

The next-generation SKAO and CTAO observatories will revolutionise dark matter searches \citep{Lelli01.2026.SKA,Regis01.2026.SKA}. Their unprecedented sensitivity will allow for much deeper probes of key targets like the Galactic Centre and dwarf spheroidal galaxies. Joint analysis strategies, leveraging distinct spectral and morphological signatures in both bands, are crucial for enhancing the signal-to-noise ratio and robustly discriminating a potential dark matter signal from complex astrophysical backgrounds. This synergy will either lead to a compelling, multi-wavelength discovery or place the most stringent constraints yet on the properties of WIMP dark matter across many channels \citep{Balazs:2015boa}.

{In addition to the GC and dwarf galaxies, galaxy clusters are also key laboratories to probe dark matter models. Indeed, as galaxy clusters are the most massive gravitationally bound structures,
embedded in dark matter halos of $\sim10^{14}~M_\odot$ or above, they are
promising targets for indirect dark matter searches \citep{Markevitch2004,Harvey2015}. In particular, self-interacting dark matter (SIDM) models predict modifications to the central density profiles, which are crucial to address the so-called {\it cusp vs core} problem in clusters \citep{Newman2011,Laporte_White2015}.  
Altogether, annihilation of dark matter can result in the production of particles including electrons and positrons that, in the presence of magnetic fields, lose energy via synchrotron radiation, observable as radio emission \citep{Storm2013,Colafrancesco2015,Chan2019,Chan2020a,Chan2020b}.
Annihilation or decay of dark matter particles in clusters 
could also contribute to the
gamma-ray and TeV flux and to flattening of the matter density of the cluster cores \citep{Perkins2006,Aharonian2009,Ando_Nagai2012,Adams2016,Hutten_Kerszberg2022,Albert2024}.} 

   \subsection{Search for Extraterrestrial Intelligence (SETI)}

While the role of radio astronomy in SETI activities is well-documented, the role of gamma-ray telescopes in this activity is less well-known. Ground-based gamma-ray telescopes utilising the air-Cherenkov technique employ very large optical reflecting mirrors (area $>100$m$^2$ in many cases) and detector cameras operating on nanosecond timescales. These features are especially useful to search for {\em technosignatures} such as very rapid transient optical flashes that might be generated by other civilisations \citep{book2,Schwartz:1961}.
    
The main example often discussed is the use of powerful laser pulses \citep{Eichler:2001}, knowing that our pulsed laser technology of today (e.g. 3\,ns 3\,MJ pulse collimated with a 10\,m mirror) could significantly outshine a solar-type star viewed from about 300\,pc \citep{Howard:2004}. The VERITAS facility has searched for pulsed optical emission from $>200$ nearby stars and some known exoplanet systems, e.g. KIC 8462852 \citep{Veritas-SETI:2016, Veritas-SETI:2023}, and has joined Breakthrough Listen multi-wavelength campaigns. VERITAS's results have so far inferred stringent limits on the fraction ($<10^{-7}$) of nearby stars hosting civilisations emitting pulses up to a $\sim$27\,hr repetition rate. The future CTAO facility will significantly improve on this.

Radio astronomy has been central to SETI efforts since its inception, starting with Project Ozma, which looked at two nearby stars~\citep{Drake1961}, to modern surveys such as Breakthrough Listen, which monitor numerous stars across multiple frequencies~\citep{2020AJ....159...86P}. For example, the SKAO precursor MeerKAT has a planned Breakthrough Listen survey to observe one million nearby stars to search for technosignatures~\citep{2021PASP..133f4502C}. Radio is particularly adept for these searches, as it excels at detecting continuous-wave signals over longer integration times and suffers minimal interstellar extinction. This allows both the analysis of any complex structure within the signal, as well as the detection of transients over longer timescales and at vast distances. The improved sensitivity and survey capabilities of SKAO will improve this even more \citep{Tremblay01.2026.SKA}.

{Recently, \citet{2025arXiv250616351L} argued that high-energy SETI could potentially test the fundamental limits of technology, and that advanced civilizations might operate near high-energy sources such as neutron stars or the accretion disks of black holes. Civilizations harvesting neutron-star radiation could potentially be detectable via gamma-ray (and radio) flux variations.}

{Furthermore, globular clusters may contain {\it sweet spots} where habitable zone planetary orbits can be stable over long timescales \citep{DiStefano_Ray2016}. Given the high stellar density compared to the Galactic disk, short interstellar distances would significantly facilitate inter-cluster communication, travel, and the establishment of outposts. This proximity reduces the risk of total extinction from localized catastrophes, making globular clusters prime SETI targets. Additionally, the high concentration of millisecond pulsars that send out high energy and radio pulses provides a natural, precise timing network that resident civilizations could utilize for spacecraft navigation using relatively small on board detectors and antennae.}

The true potential, however, lies in coordinated multi-frequency campaigns; in particular, optical-radio synergies offer particular advantages \citep{Bagchi01.2026.SKA}. Together, radio and optical can combine both the exceptional time resolution and complex wave signature analysis to help discriminate between astrophysical and article origins. For example, natural phenomena such as stellar flares typically produce correlated optical and radio emissions with predictable spectral characteristics, whereas artificial signals may exhibit uncorrelated or structured patterns across different frequencies. 

This synergy provides unique opportunities for detecting complex signals from potentially advanced civilisations that may employ multiple transmission strategies simultaneously. Coordinated observations would allow both the identification of any complex structure within the signal due to the radio's larger integration time, as well as precise timing references for any optical counterparts. This would allow the detection of complex signals that would be invisible to single-frequency searches.

\section{Conclusions}
In this work, we have presented and discussed the potential of exploiting SKA, in synergy with ongoing and forthcoming gamma-ray facilities, to observe a wide variety of high-energy astrophysical sources. They include variable and transient targets (Sect.~\ref{sec:variable_and_transients}) and steady-state and extended sources (Sect.~\ref{sec:steady_state_sources}). The synergy between SKAO and gamma-ray observatories and telescopes will potentially address still open questions of modern astrophysics, thus helping to improve our understanding of a number of key topics. They include (i) non-thermal particle acceleration and high-energy emission mechanism phenomena; (ii) multi-messenger astrophysical processes including the production of neutrinos, gravitational waves from compact sources, and cosmic ray acceleration; (iii) the nature of dark matter and physics beyond the standard model.

    \bibliographystyle{abbrvnat-maxbibnames4}  

    \bibliography{main_rev2} 
    
    \end{document}